\documentclass[conference]{IEEEtran}
\IEEEoverridecommandlockouts

\usepackage{amsmath,amssymb,amsfonts}
\usepackage{algorithmic}
\usepackage{graphicx}
\usepackage{textcomp}
\usepackage{xcolor}
\usepackage{comment}
\usepackage{tablefootnote}

\usepackage{multibib}

\usepackage{xparse}
\newcites{review}{GPT Game Literature}

\def\BibTeX{{\rm B\kern-.05em{\sc i\kern-.025em b}\kern-.08em
    T\kern-.1667em\lower.7ex\hbox{E}\kern-.125emX}}
\begin{document}

\title{GPT for Games: A Scoping Review (2020-2023)\\
}

\author{\IEEEauthorblockN{Daijin Yang}
\IEEEauthorblockA{\textit{College of Art, Media, and Design} \\
\textit{Northeastern University}\\
Boston, United States \\
yang.dai@northeastern.edu}
\and
\IEEEauthorblockN{Erica Kleinman}
\IEEEauthorblockA{\textit{College of Art, Media, and Design} \\
\textit{Northeastern University}\\
Boston, United States \\
e.kleinman@northeastern.edu}
\and
\IEEEauthorblockN{Casper Harteveld}
\IEEEauthorblockA{\textit{College of Art, Media, and Design} \\
\textit{Northeastern University}\\
Boston, United States \\
c.harteveld@northeastern.edu}
}

\maketitle

\begin{abstract}


This paper introduces a scoping review of 55 articles to explore GPT's potential for games, offering researchers a comprehensive understanding of the current applications and identifying both emerging trends and unexplored areas. We identify five key applications of GPT in current game research: procedural content generation, mixed-initiative game design, mixed-initiative gameplay, playing games, and game user research. Drawing from insights in each of these application areas, we propose directions for future research in each one. This review aims to lay the groundwork by illustrating the state of the art for innovative GPT applications in games, promising to enrich game development and enhance player experiences with cutting-edge AI innovations.

\end{abstract}


\section{Introduction}
The advent of large language models (LLMs) has transformed human-AI interaction. The powerful learning capabilities of LLMs enable them to perform a wide range of tasks through context-specific fine-tuning or prompt engineering, altering the traditional interaction mode that required retraining the entire model~\cite{training}. Among LLMs, the GPT series, including GPT-2~\cite{GPT2}, GPT-3~\cite{GPT3}, ChatGPT, GPT-3.5, and GPT-4~\cite{GPT4}, has garnered widespread attention in both the academic and industrial communities~\cite{GPTsurvey} due to its leading position in both publishing time~\cite{GPTsurvey, LLMtime} and performance~\cite{compare1, compare2}. It has shown remarkable flexibility and utility across a wide range of applications~\cite{GPTsurvey}, including information extraction~\cite{entityextraction}, question-answering~\cite{QA1,QA2}, text generation~\cite{textgeneration}, programming tasks~\cite{programming}, and creativity support~\cite{aiasactive,creativitysupport}. 
This breadth of functionality highlights the multifaceted nature of GPT and underscores its potential to transform various aspects of game development and interaction.

Research on GPT for games has already begun. Leveraging GPT's powerful natural language processing capabilities, it is used extensively for generating and processing game-related text, such as character dialogue~\citereview{14} or entire stories~\citereview{7}. In addition to narrative creation, it has also been used to generate design ideas for board games~\citereview{37}, create music~\citereview{17}, organize 3D scenes~\citereview{50}, and host Dungeons and Dragons~\citereview{20}, a classic tabletop role-playing game (TTRPG).

Despite this, the exploration of GPT for games is still in an early stage. At this point, with signs of rapid growth in the near future, a scoping review would be beneficial, enabling researchers to comprehensively grasp the current state of the art of GPT for games and identify both trends and gaps. Such a review would provide essential support by highlighting opportunities for future research and best practices.

In this work, we provide such a review, focusing on the current use cases of GPT in games research. After a systematic search and screening process, 55 articles were identified for review. By open-coding the papers, and categorizing the codes, we identified five prominent use cases for GPT in contemporary games research: procedural content generation, mixed-initiative game design, mixed-initiative gameplay, playing games, and game user research. 
Based on the insights gained from examining each of the application areas, we suggest future research directions for each category identified.


\section{Methodology}

In our review, we employed the keywords ``game'' and ``GPT'' to search the following databases for articles relevant to our research topic: ACM Digital Library, IEEE Explore Digital Library, Springer, and AAAI.
For the first three, we conducted searches directly using the library search engines provided on their respective websites.
For the AAAI papers, due to the lack of an advanced search feature in AAAI's online library, we substituted Google Scholar as the search engine, limiting the search results to articles sourced from aaai.org. We searched full texts in all databases.

The preliminary search results (\textit{n} = 2,098) using the aforementioned keywords were as follows: 877 articles from ACM, 903 from IEEE Explore, 62 from AAAI, and 256 from Springer. We then screened the preliminary search results based on the following criteria:
\begin{itemize}
    \item [1)] The articles must involve the study of a system related to digital or analog games.
    \item [2)] The articles must report interactions of either authors or their participants with any version of GPT.
    \item [3)]The articles were published before January 1, 2024.
\end{itemize}
After reviewing the titles and abstracts, 67 papers directly related to games and mentioning GPT advanced to the next stage. A large number of articles were excluded at this stage because 'game' is a broadly used term, not solely referring to the context of games discussed in this paper. 
Upon a detailed examination of the 67 papers, 4 articles were excluded based on the first criterion, and 8 articles were excluded based on the second criterion.

Ultimately, 55 articles were included in the final literature synthesis.
To track the trends, one researcher compiled the publication dates of articles and the GPT models studied within them. When an article utilized multiple GPT models, the model that was the primary focus of the research was counted. If multiple models played identical roles in the study, the most recent model was counted.
We consider ChatGPT and GPT-3.5 as two different models for two reasons. First, ChatGPT can be powered by GPT-4, and most research using ChatGPT did not specify which version they used. Second, articles using ChatGPT were based only on web interactions with OpenAI's site, while those using GPT-3.5 interacted through API.
To extract the use cases of GPT for games, one researcher open-coded~\cite{opencoding} the papers, focusing on identifying what content has been generated by GPT, how the content has been used in a game-related system, and how users interacted with GPT and/or the generated content. Five major categories emerged from this process. For each category, the researcher further sorted the papers into sub-categories based on what content had been generated and how the content had been used. The categorizations and papers were then discussed and refined by two other researchers, leading to the final grouping.




\section{Results}

\subsection{General Research Trends of GPT for Games}

\begin{figure}[htpb]
\begin{minipage}[b]{1.0\linewidth}
  \centering
  \centerline{\includegraphics[width=9cm]{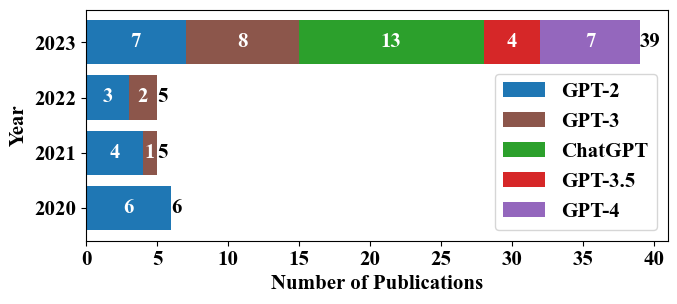}}
\end{minipage}
\caption{The General Trends of GPT for Games.}
\label{fig:trend}
\end{figure}

As shown in Figure~\ref{fig:trend}, the first paper we retrieved in the database was published in 2020, following the release of GPT-2 in 2018~\cite{GPT2}. The number of publications in 2021 and 2022 demonstrates an increasing interest in the new GPT-3 model. Following the release of ChatGPT and GPT-4, there was a substantial increase in publications in 2023. 
This sharply rising trend coincides with the performance enhancements of the GPT models~\cite{gpt2vsgpt3} and suggests that further research on the latest GPT models is to be expected.

\subsection{Use Cases of GPT for Games}
\begin{figure}[htpb]
\begin{minipage}[b]{1.0\linewidth}
  \centering
  \centerline{\includegraphics[width=9cm]{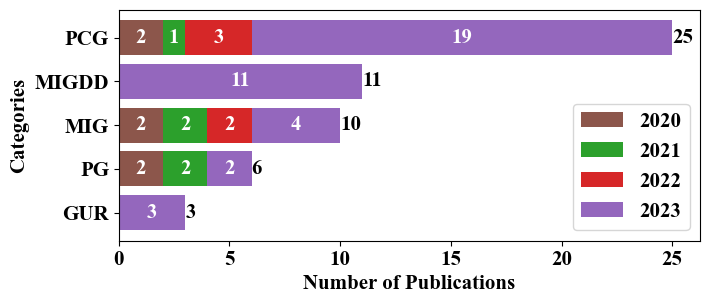}}
\end{minipage}
\caption{The General Trends of Each Category.}
\label{fig:trendyear}
\end{figure}

As shown in Figure~\ref{fig:trendyear}, 25 papers focused on using GPT for procedural content generation (PCG). Eleven papers focused on using GPT for mixed-initiative game design and development (MIGDD). Ten papers used GPT for mixed-initiative gameplay (MIG). Six papers used GPT to play games (PG). Finally, three papers used GPT for game user research (GUR). In this section we report in terms of these use cases.


\subsubsection{GPT for Procedural Content Generation (PCG)}
In this category, GPT generated game content, at the time of gameplay, based on constraints set by designers at the time of game development. No iteration or human input was involved in that generation~\cite{pcgsurvey}.
Before 2023, only the GPT-2 model had been used for this category (\textit{n} = 6). In 2023, researchers published more studies utilizing subsequent versions of GPT models (\textit{n} = 13). Nevertheless, GPT-2 continued to be a popular choice for PCG in 2023, especially for story (\textit{n} = 2), quest (\textit{n} = 2), and level generation (\textit{n} = 2).

\textbf{GPT has been used for story generation (\textit{n} = 14,~\citereview{18, 34,38,15,46,7,23,28,22,14,54,20,25,26}).} Due to GPT's strong natural language processing abilities, the majority of the work focused on procedural story generation. Compared to advanced GPT versions, using GPT-2 (\textit{n} = 5) for story generation demanded additional efforts, including fine-tuning and employing supplementary techniques. Fine-tuning was necessary for all GPT-2 related studies to tailor their output more closely to the specific context, requiring a dataset closely aligned with the research objectives~\citereview{46}. For example, in~\citereview{26}, GPT-2 was fine-tuned with fantasy stories to craft narratives for a Dungeons and Dragons (D\&D) game. To address GPT-2's limitations in language processing and memory capacity, further techniques were introduced. For instance,~\citereview{7} utilized a knowledge graph with GPT to produce consistent stories. Meanwhile, in~\citereview{26}, BertScore~\cite{bertscore} was employed to evaluate GPT-2's outputs, ensuring that only the highest-quality texts were presented to players, thus mitigating the issue of low-quality content. 
GPT-3, ChatGPT, GPT-3.5, and GPT-4 demonstrate superior in-context capabilities~\cite{GPT3} for story generation compared to GPT-2, where only proper prompt engineering is necessary to manage the task. 
Prompts for story generation can be simple and only specify the tasks GPT should perform. For instance, in~\citereview{38}, ChatGPT was asked to generate a story for a board game by ``\textit{(Player) landed on the same tile. Generate a battle story for the players.}''

\textbf{GPT has been used for quest generation (\textit{n} = 4,~\citereview{27,41,47,31}).}
Unlike a story, a quest also encompasses a set of actions that the player must perform~\cite{quest}. This distinction makes the approach to generating quests with GPT different from that of generating stories.
GPT-2 was the only model that was used for quest generation. The common techniques employed across these investigations include fine-tuning the GPT-2 model on annotated datasets specific to RPG quests and using prompt engineering to guide the model's output toward generating quests and dialogues that fit the structure and style of existing game content. For example, in~\citereview{27}, the authors explored the use of GPT-2 for generating quests in a ``World of Warcraft'' setting. They fine-tuned it with a corpus of quest texts, resulting in the model's ability to produce quests that not only matched the thematic elements of the game but also maintained logical coherence and creativity. Additionally, in~\citereview{41}, the use of a knowledge graph for quest generation was explored, offering a method to integrate game-world elements like locations, NPCs, and items into quests. This approach ensured quests were contextually relevant and personalized, enhancing player immersion by tailoring content to the game's lore and the player's actions.
 
\textbf{GPT has been used for level generation (\textit{n} = 3,~\citereview{45,36,43}).}
GPT-2 and GPT-3 models can effectively be fine-tuned with text-encoded designs for game level generation. For example, in~\citereview{45}, MarioGPT, a fine-tuned GPT-2 model, was developed for creating Super Mario Bros levels. This model leverages text representations of level elements, allowing for the generation of levels through natural language prompts. This method provided precise control over game design elements like enemy distribution and landscape features, illustrating the effectiveness of text-based representations in enhancing procedural content generation techniques.

\textbf{GPT has been used for character generation (\textit{n} = 3,~\citereview{1,21,25}).} 
ChatGPT and GPT-3 were used to generate attributes for characters. For instance, in~\citereview{25}, ChatGPT was used to generate race, alignment, gender, ability scores, and equipment for a character in a D\&D game. In \citereview{1}, GPT-3 was used to generate energy, speed, and control parameters for a virtual coach in a fitness game. Notably, the detailed methods of designing such systems were not reported in all cases.

\textbf{GPT for other PCG Uses (\textit{n} = 2,~\citereview{17,5}).}
Our review revealed two other instances of GPT being used to generate content for games, in the first~\citereview{17}, it was used to generate music, and in the second~\citereview{5}, it was used to generate real-time commentary. \citereview{17} presented Bardo Composer, a system designed to generate background music for tabletop role-playing games, using a speech recognition system to convert player speech into text. This text was then classified according to a model of emotion, and the system generates music pieces conveying the desired emotion using a novel Stochastic Bi-Objective Beam Search algorithm~\cite{SBBS}. In~\citereview{5}, the authors introduced a prompt engineering methodology using GPT-3.5 for generating dynamic commentary in fighting games, emphasizing the impact of prompt design on output quality. The work highlights the preference for simpler prompts by users for more engaging commentary.

\subsubsection{GPT for Mixed-Initiative Game Design and Development}
Similar to the previous category, GPT was used to generate content for design and development purposes, but unlike the previous category, the generation was an iterative process~\cite{mixedinitiativeinterface}, where the designer worked collaboratively with GPT over several attempts to generate content that was then installed within the game.
All papers in this category were published in 2023 and ChatGPT emerged as the most popular model (\textit{n} = 5).

\textbf{GPT can aid in creating game scenarios (\textit{n} = 5,~\citereview{39,30, 49,50,40}).} 
The most direct application is using GPT to assist with text writing in scenarios.
For example, in~\cite{39}, the authors leveraged ChatGPT as a collaborative writing tool to generate dialogue for their game. They provided it with the setting of their game world and asked it to articulate pros and cons for a particular issue to use as character dialogue in their game. 
In addition to story co-writing, GPT can aid in designing scenes. In~\citereview{40}, the authors designed an educational matching game. The game requires players to read an article and match the concepts mentioned in the article with their descriptions. Through multiple interactions, the author prompted GPT to extract the concepts from the reading materials and generate detailed descriptions for these concepts, which were then placed into the game scenario. 
Beyond text outputs, GPT can also aid in generating character expressions and 3D game scenes.
In~\citereview{49}, it was employed to create dialogues from initial prompts for selected characters. Moreover, GPT could select suitable pre-designed expressions and movements for characters based on the dialogue. In~\citereview{50}, it was capable of arranging 3D furniture in a Unity scene according to the input prompt. The authors in~\citereview{49} and~\citereview{50} utilized GPT to process and generate data in JSON format for system implementation. For instance, in~\citereview{50}, GPT-4 was prompted to generate a JSON file containing keys such as object name, X, Y, and Z positions, and facing direction. The authors provided detailed definitions for each key in the prompt.

\textbf{GPT can help to design game mechanics and rules (\textit{n} = 5,~\citereview{2,33,11,37,12}).} 
In~\citereview{2}, the authors fine-tuned GPT-2 to generate game features based on short game descriptions. The generated features, they argue, can be used for conceptual game design.
In addition to generating simple features, more work involves integrating GPT with other frameworks to create detailed game mechanics and rules.
In~\cite{33}, the authors leveraged GPT to design gamified mechanics for debugging and troubleshooting activities in software engineering.
They requested it to generate outputs according to the gamification design framework provided by Dal Sasso et al.~\cite{gamify}, through prompt engineering.
In~\citereview{11}, the authors outlined a method using the evolutionary algorithm for GPT to create playable board games. In the algorithm, GPT first generated a set of random game design proposals as the initial population. Players then scored these designs. After obtaining the scores, GPT merged two designs selected through tournament selection~\cite{tournament} to create a new proposal and altered certain parts of the new proposal to introduce mutations. Subsequently, the new proposal was added back into the population, which was then scored by players, entering a cycle. After 30 cycles, a total of 40 game design proposals were produced.
In~\citereview{37}, GPT was guided to follow the Design Sprint framework~\cite{designsprint} for board game design. The Design Sprint framework consists of five phases: understand, diverge, converge, prototype and test. GPT was prompted to perform the following tasks in sequence: research existing board games (understand and diverge), create a board game with constraints (converge and prototype), reflect on game design (test), and follow the mixed-initiative interaction to update the game based on reflections and player feedback multiple times(iterate).

\textbf{GPT can assist with programming tasks (\textit{n} = 2,~\citereview{12,51}).} In~\citereview{12}, the author reported participants in a game jam utilized GPT to program their entire game. The detailed interactions were not reported in the paper.
In~\citereview{51}, the authors aimed to enrich the text embedding space of a text-to-image model designed for creating pixel art game sprites. To achieve this, they input the existing dataset labels into GPT-4 along with a custom prompt, instructing it to generate alternate versions of each label using simple language.


\subsubsection{GPT for Mixed-initiative Gameplay}
Unlike the previous section, which leverages GPT to aid in design, this section focuses on GPT as an aid during play.
In this category, the number of publications saw a slight increase to 4 in 2023, compared to 2 in each year before that. Additionally, with the release of new GPT versions, research in this category tends to utilize the latest GPT models for the study.

Our review found 8 examples where GPT supported mixed-initiative story creation for narrative games (\textit{n} = 8,~\citereview{9,44,16,19,48,35,24,8}).
The interaction entails players and GPT alternating turns to add to a story, which can include contributing sentences and paragraphs within the narrative. 
The interaction began with a starting prompt for GPT or players to initiate the story. GPT would use the previously created story as its input to generate new content. For example, in~\citereview{8}, GPT-3 was given the prompt ``\textit{This is the start of an ancient Middle Eastern mystery story.}'' at the beginning. Then, players and GPT-3 would take turns creating this story in the game.
Besides simple prompt engineering, GPT-2 and GPT-3 can be fine-tuned to generate texts with special requirements.
For instance, in~\citereview{19}, the authors aimed to examine the impact of a bot's creativity on children's creativity skills within a storytelling context.
They created different GPT-3 models that had varying levels of creativity by fine-tuning them with various stories acquired from their previously conducted studies. These models were then used in a mixed-initiative storytelling game, where participants and the GPT-3 models took turns adding a sentence to a fantasy story until the participants inputted 'end'. The GPT-3 models would generate their sentences based on what the participants had previously written.
Additionally, GPT can evaluate human-written texts, filtering out low-quality inputs to maintain gameplay and providing players with feedback and suggestions. 
For example,~\citereview{8} used an additional GPT-4 bot to evaluate human input, where the characters in the game would refuse to advance the storyline if the human input was deemed irrelevant to the context.

Moreover, we found 2 examples where GPT supported storytelling for dungeon masters (\textit{n} = 2,~\citereview{10,53}).~\citereview{10} and~\citereview{53} demonstrated the application of GPT-3 in assisting Dungeon Masters (DMs) for Dungeons \& Dragons (D\&D) and Game Masters (GMs) for other tabletop role-playing games (TTRPGs). They explored how GPT-3 and ChatGPT can offer creative support by generating enemy descriptions and configurations, summarizing the current game situations, and brainstorming narratives.
For example, in~\citereview{10}, GPT-3 has been used to summarize and simplify the descriptions of various monsters in the D\&D rulebook, aiding DMs in quickly understanding them. Additionally, ChatGPT has been utilized to assist DMs in brainstorming stories for encounters. The authors designed separate interfaces for each function and crafted distinct prompts to enable GPT to perform these tasks.

\subsubsection{GPT Can Play Games}

Works in this category explored how GPT can autonomously play games or serve as a virtual opponent. Most of the work (\textit{n} = 4) in this category utilized GPT-2. Nevertheless, newer models, such as GPT-3.5 and ChatGPT, have also been employed in the latest research.
GPT-2 demonstrated a strong capability in word-guessing games. 
This includes gameplay like deducing a word based on its description~\citereview{4}, an attacker compelling a defender to utter a specific word while the defender attempts to identify it~\citereview{3}, and engaging in board games where the objective is to guess words through associative clues~\citereview{55}. 
In these studies,~\citereview{3} and~\citereview{55} did not fine-tune GPT-2 but directly utilized its word prediction capabilities~\cite{GPT2}, using it to only generate the next word based on the input as the result of a guessing game. In contrast, in~\cite{4}, the game took place in a natural language conversation setting, requiring GPT's output to be complete sentences. The authors fine-tuned GPT using data collected from Reddit to ensure its output met the context requirements.
Besides word-guessing games, GPT-3.5 and ChatGPT were prompted to play a debate game~\citereview{13} and a game of writing stories with keywords~\citereview{42}. 
In study {13}, the authors used prompt engineering to have GPT-3.5 simulate students of varying levels, from kindergarten to university. They provided it with custom prompts for various debate stages, including responding, asking questions, and making arguments, allowing it to join the discussion. \citereview{42} utilized prompts similar to those used for story generation with GPT, simply asking it to write a story based on given keywords.
In addition to text-based games, \citereview{52} presented a novel approach to generating plausible strategic moves in the game of Go by GPT-2. By training it on a dataset of Go game records in Smart Game Format (SGF)~\cite{SGF}, the model learned to mimic the strategic styles of Go champions, producing valid and novel game strategies.

\subsubsection{GPT for Game User Research}

We found 3 examples where GPT has been used to assist game user research. 
In~\citereview{29}, the authors explored the potential of GPT-3 to analyze game reviews for enhancing game design. The authors focused on how AI can provide insights into player experiences and preferences by prompting GPT-3 to answer questions from the Player Experience Inventory (PXI)~\cite{pxi} based on players' reviews of a game.
In another example,~\citereview{6} explored the use of ChatGPT for categorizing audience comments during live game streaming to increase engagement and stream value. ChatGPT was prompted to classify audience members into five predefined categories, based on their comments and the live streamer's current game status.
Additionally, in~\citereview{32}, the authors investigated and compared human and generative AI (GPT-4 and ChatGPT) participants' answers to interview questions about voice interaction across three scenarios: game-to-player, player-to-player, and player-to-game interactions.

\section{Discussion}



Although newer and larger GPT models demonstrate strong capabilities in games without the need for fine-tuning or additional programming techniques, smaller models such as GPT-2 still remain valuable for further research and development. They have lower operational costs, fewer ethical constraints, and more flexibility in modifications. Moreover, they have the potential to surpass the general capabilities of larger GPT models with specific modifications.
In other domains, including healthcare~\cite{healthcare}, legal~\cite{legal}, and finance~\cite{finance}, small GPT models trained with domain-specific knowledge have shown superior performance compared to larger general GPT models.
Generating game-related texts can also be domain-specific, involving different terminology in gameplay~\citereview{10}, game design~\cite{mda}, and game analytics~\cite{terms}.
Future research could benefit from investigating the customization of smaller GPT models for gaming applications and comparing their effectiveness with that of generally applied larger GPT models.

Due to GPT's powerful text generation capabilities, it is unsurprising that generating game-related text has garnered the most attention. Compared to story generation, other tasks can be relatively more complex, and GPT-2 was mostly used for these studies. Quest generation requires generating context-appropriate text but also understanding the relationships between entities in the narrative~\cite{41}. Level, scene, character, and music generation require generating output in a special format, as well as extracting sentiments~\citereview{17,49}, planning the space~\citereview{50}, and considering gameplay~\cite{questsdesign,leveldesign} and player experience aspects~\cite{leveldesign,musicdesign}.
Unlike GPT-2, the newer GPT models have more powerful abilities that can support these tasks, such as entity extraction capabilities~\cite{entityextraction} and spatial and situational understanding abilities~\cite{spacial}. 
However, exploration into applying new models in non-story generation tasks remains quite limited.
Future research could consider using the latest GPT models for these tasks, exploring prompt engineering, and integrating generated content into actual games to explore player feedback.

Additionally, most research on GPT for mixed-initiative game design has focused on single aspects of the design and development process. However, game design and development involve multiple, interconnected tasks~\cite{mda,gamedesignmulti}. Future studies could integrate various GPT models to create a comprehensive game development support system, 
potentially improving context understanding for GPT~\citereview{33}, enhancing suggestion relevance, and increasing efficiency. This approach has been explored in software development~\cite{chatdev}, where GPTs fulfilled roles like designers and developers, working together to produce a usable application, though the focus was on code generation. A similar, but largely unexplored, integration has been reported for game development~\cite{gamegpt}. Further research is encouraged to combine GPT's functionalities for game design, 
investigate prompt engineering techniques like creating chains that are incorporated across multiple GPT models~\cite{promptengineeringreview},
explore human-computer interaction models, and provide detailed reports on the interaction experiences and outcomes.

We also found that most work in mixed-initiative gameplay involves similar game mechanics with a lack of user experience reports. Interactions between players and GPT typically involve simple story continuations without complex mechanics like puzzle solving. Research on GPT's use in assisting gameplay is also scarce and mostly focused on tabletop games. GPT has the potential to provide personalized tutorial support~\cite{tutorial} or story summaries for different levels of players.
Future research could explore integrating GPT into games with complex mechanics, examine various player-GPT interaction modes, report on user experiences, and provide design suggestions for using GPT in games.

Beyond generation tasks, an in-depth exploration into GPT playing games is still lacking.
It is undeniable that past research on AI playing games~\cite{deepblue,alphago,jaderberg2019human,vinyals2017starcraft} has advanced the field of artificial intelligence~\citereview{3}.
For instance, the iterative development from AlphaGo~\cite{alphago} to MuZero~\cite{muzero} has furthered research in reinforcement learning AI. Presently, investigations into GPT's involvement in games predominantly revolve around language-based games, which may encompass logic reasoning, scenario extraction, and comprehension~\citereview{55,3}. However, games beyond the scope of language pose additional demands on GPT's abilities, such as the need for AI to accurately understand rules, develop complex strategies, and decipher hidden game states~\cite{hearthstone}. Future studies could broaden their horizons by employing a diverse array of games to probe the capabilities and limitations of newer GPT models.

Finally, the application of GPT in the field of game user research is still very limited. Considering GPT's powerful text processing capabilities, future research could explore using GPT to replace and compare with other AIs or humans in processing text within game user research. For example, GPT could be used to predict toxic speech as a replacement or comparison to the Random Forest model~\cite{toxic}, extract player emotions from comments as a replacement or comparison to other natural language processing techniques~\cite{sentimentextract}, or in analyzing qualitative user experience data~\cite{coding}.

In the process of conducting our scoping review, we noticed that few articles deeply discuss the limitations and ethical problems associated with GPT. We encourage future research to more thoroughly explore issues related to hallucination, privacy, security, fairness, and prompt attacks in the context of games. 

\section{Limitation}
Our research inevitably faces limitations in capturing the ongoing or yet-to-be peer-reviewed studies. This rapid development means that new use cases for GPT in gaming are emerging frequently, which may not be fully reflected in our current survey. 
Moreover, although the databases we chose include mainstream conferences and journals in the field of gaming, we acknowledge that the databases we did not select may also contain topics relevant to our paper. 
Additionally, our focus has been primarily on academic publications, excluding a review of commercial games that integrate GPT technologies. The gaming industry often pioneers innovative uses of such technologies, offering practical insights and challenges that are crucial for a holistic understanding. 
Future work will aim to bridge these gaps by incorporating a broader spectrum of sources, including industry-based applications and the latest research findings, to offer a more comprehensive view of the capabilities and limitations of GPT in gaming. 

\section{Conclusion}
We conducted a scoping review of 55 articles related to the application of GPT for games. Through this review, we identified various uses of GPT within game research. The applications include using GPT for procedural game content generation, employing GPT in mixed-initiative game design and gameplay processes, leveraging GPT to autonomously play games, and using GPT for game user research.
Based on our findings, we recommend future studies to concentrate on experimenting with game-specific GPT models, employing the latest GPT versions for procedural content generation tasks beyond story creation, developing game design tools that utilize multiple GPT models for diverse tasks, incorporating GPT into complex games to explore player experience and GPT's potential in gameplay, and leveraging GPT as a substitute for traditional text analysis methods in game user research.
With this work, we aim to set the groundwork for more innovative applications of GPT in gaming, thereby enhancing both game development and gameplay experiences through advanced AI techniques.

\end{document}